\documentstyle[aps,prb,eqsecnum,epsf]{revtex}
\begin{document}
\title{Bias-voltage induced phase-transition in bilayer quantum Hall ferromagnets}
\author{Yogesh N. Joglekar and Allan H. MacDonald}
\address{Department of Physics, University of Texas at Austin, Austin, Texas 78712, \\ 
Department of Physics, Indiana University, Bloomington, Indiana 47405. }
\maketitle
\begin{abstract}
We consider bilayer quantum Hall systems at total filling factor $\nu=1$ in presence of a bias voltage 
$\Delta_v$ which leads to different filling factors in each layer. We use auxiliary field functional 
integral approach to study mean-field solutions and collective excitations around them. We find that 
at large layer separation, the collective excitations soften at a finite wave vector leading to the 
collapse of quasiparticle gap. Our calculations predict that as the bias voltage is increased, bilayer 
systems undergo a phase transition from a compressible state to a $\nu=1$ phase-coherent state 
{\it with charge imbalance}. We present simple analytical expressions for bias-dependent renormalized 
charge imbalance and pseudospin stiffness which are sensitive to the softening of collective modes.
\end{abstract}     

%------------------------------------------------------------------------------------------------------%

\section{Introduction}
\label{sec: intro}

Bilayer quantum Hall systems consist of a pair of two-dimensional (2D) electron gases separated by a 
distance $d$ which is comparable to the inter-particle spacing within each layer. In a strong magnetic 
field perpendicular to the layers, the kinetic energy of electrons is quantized and the macroscopic 
degeneracy of Landau levels accommodates all electrons in the lowest Landau level, thereby enhancing 
the role of interactions in determining the physical properties of the system. Bilayer systems at 
total filling factor $\nu=1$ in the absence of bias voltage have broken symmetry ground states that 
can be regarded as either easy-plane ferromagnets or as exitonic superfluids, and have been 
intensively investigated over the past 
decade.~\cite{qhereviews,haf,apb,wenzee,ezawa,sqm,km,ky,spi1,balents,stern,fogler,dns,spi2,kynew,ya3} 
At zero temperature the Zeeman coupling of electron spins with the strong magnetic field polarizes all 
spins along the field. Therefore, the only dynamical degrees of freedom are the orbit centers within a 
Landau level and the discrete layer index which we describe using a pseudospin label with ``up'' 
denoting a state localized in the top layer and ``down'' denoting a state localized in the bottom 
layer. As the layer separation $d$ is increased, balanced bilayer systems undergo a quantum phase 
transition from a phase-coherent $\nu=1$ quantum Hall state (where each layer has a filling factor 
$\nu=1/2$) to a disordered compressible state at a critical layer separation $d_{cr}$ which increases 
with interlayer tunneling amplitude $\Delta_t$.~\cite{apb,sqm,spi1,spi2} The $\nu=1$ quantum Hall 
effect observed for $d<d_{cr}$ is due solely to strong interlayer correlations, since each layer has a 
filling factor $\nu=1/2$, a compressible-state fraction for an isolated layer. When the bias-voltage 
is zero, the mean-field many-body ground state has all electrons occupying symmetric bilayer Landau 
level states, or equivalently all pseudospins pointing along the positive $x$-axis. At finite $d$, 
because of the difference between the intralayer Coulomb interaction $V_A(\vec{q})=2\pi e^2/\epsilon q$
 and interlayer Coulomb interaction $V_E(\vec{q})=V_A(\vec{q})e^{-qd}$, the pseudospin polarization 
$m_x$ does not commute with the interaction. Therefore the fully polarized mean-field state is not an 
eigenstate of the Hamiltonian and in particular, it is not the exact ground state. 

In this paper we study a bilayer system at total filling factor $\nu=1$ {\it in the presence of a 
bias voltage} $\Delta_v$. In this case the two layers have different filling factors, $\nu_T$ (top) 
and $\nu_B$ (bottom) respectively, and the dimensionless charge-imbalance $m_z=(\nu_T-\nu_B)$ is 
proportional to the applied bias voltage. The system takes advantage of the interlayer Coulomb 
exchange interaction by spontaneously generating a local alignment of pseudospins which leads to 
interlayer phase coherence. As a first approximation, we use the Hartree-Fock approximation 
to obtain the mean-field quasiparticles, and to estimate the charge imbalance $m_z$ and the pseudospin 
polarization $m_x$. We study the stability of Hartree-Fock solutions by considering the spectrum of 
collective excitations around them. The collective excitations in these systems are properly described 
only when fluctuations in both exchange and direct particle-hole channels are included~\cite{haf} and 
the auxiliary field functional integral approach which we use here naturally takes these into 
account.~\cite{kerman,ya1} We find that increasing bias-voltage stiffens collective excitations at 
finite wave-vector and reduces the importance of quantum fluctuations. We use the functional integral 
approach to evaluate the ground state energy and its dependence on external fields, which, in turn, 
allow us to obtain simple results for the bias-dependence of the renormalized charge-imbalance $m_z$, 
the anisotropy mass $\beta$ (or equivalently the interlayer capacitance $\beta^{-1}$) and the 
pseudospin stiffness $\rho_s$. 

The effect of bias-voltage on interlayer phase-coherence has been studied previously by L. Brey 
{\it et al.}~\cite{brey} Our results, to the extent that the papers overlap, are in accord with the 
earlier work. Here we solve the mean-field equations analytically, and are able to use the analytic 
expressions for collective mode energies to 
develop a theory of quantum fluctuations in the mean-field ground state. In addition, we emphasize the 
experimental implications of the compressible-to-incompressible state phase-transition induced by 
applying a bias voltage.

Ours is a weak coupling theory which associates the transition between phase coherent and incoherent 
states with an instability to a quantum-disordered pseudospin density wave state.~\cite{apb} The 
predictions we make here for the bias voltage dependence of coherent state properties follow directly 
and in a physically transparent way from the weak coupling theory of coherent state properties. Since 
other pictures exist, in particular for the nature of the large-layer-separation 
state,~\cite{dns,kynew} comparison of our predictions with experiment should prove valuable. 

The paper is organized as follows. Section~\ref{sec: mf} presents the Hartree-Fock analysis and the 
calculation of the spectrum of collective excitations using the functional integral approach. In 
Sec.~\ref{sec: phase} we present the principle result of this paper, namely the phase-diagram of a 
bilayer system in the parameter space $(d,\Delta_v)$, and discuss its experimental implications. In 
Sec.~\ref{sec: renorm} we present simple analytical results for parameters of the effective field 
theory Hamiltonian which describes the phase-coherent, charge-unbalanced bilayer system. We show that 
these parameters are substantially renormalized from their mean-field values as the phase boundary 
between the compressible and the incompressible regions is approached. We conclude with a brief 
discussion in Sec.~\ref{sec: summary}.

%------------------------------------------------------------------------------------------------------%

\section{Mean-field Theory and Collective Excitations}
\label{sec: mf}

Let us consider a double-layer system at total filling factor $\nu=1$ with interlayer tunneling 
$\Delta_{t}$ and bias voltage $\Delta_v$. In the strong-field limit, the Hamiltonian in 
second-quantized form is 
\begin{eqnarray}
\label{eq: mf1}
\hat{H_0}-\mu\hat{N}& = &-\sum_{k,\sigma}\left[
\frac{\Delta_t}{2}c^{\dagger}_{k\sigma}c_{k\overline{\sigma}}+
\frac{\Delta_v}{2}\sigma c^{\dagger}_{k\sigma} c_{k\sigma} \right], \\
\label{eq: mf2}
\hat{V} & = & \frac{1}{2}\sum_{k_i,\sigma_i}
\langle k_1\sigma_1 k_2\sigma_2| V_0 + \sigma_1\sigma_2 V_x|k_3\sigma_1 k_4\sigma_2\rangle 
c^{\dagger}_{k_1\sigma_1}c^{\dagger}_{k_2\sigma_2}c_{k_4\sigma_2}c_{k_3\sigma_1}.
\end{eqnarray}
where $\sigma=\uparrow=-\overline{\sigma}$ is denotes the state localized in the top layer, the $k_i$ 
are $y$-components of the canonical momenta which are good quantum numbers of the single-particle 
Hamiltonian in the Landau gauge ${\bf A}=(0,Bx,0)$, and $V_{0,x}=(V_A\pm V_E)/2$ are linear 
combinations of intralayer and interlayer Coulomb interactions. The constraint $V_0(\vec{0})=0$ 
reflects overall charge neutrality, whereas $V_x(\vec{0})=\pi e^2 d/\epsilon$ indicates the capacitive 
energy cost associated with uniform charge imbalance. In the absence of bias-voltage, it is known that 
the splitting between the symmetric and the antisymmetric state energies is 
exchange-enhanced,~\cite{haf,apb} $\Delta_{SAS}=\Delta_t+\Delta_{sb}$, where ($\lambda=0,x,A,E$) 
\begin{eqnarray}
\label{eq: mf3}
\Delta_{sb} & = & \Gamma_E(0), \\
\label{eq: mf4}
\Gamma_{\lambda}(\vec{q})& = &\frac{1}{A}\sum_{\vec{p}}V_{\lambda}(\vec{p})e^{-p^2l^2/2}
e^{i\hat{z}\cdot(\vec{q}\times\vec{p})l^2},
\end{eqnarray} 
and $l$ is the magnetic length. The $\Gamma_\lambda(\vec{q})$ are effective exchange interactions 
between an electron and a hole~\cite{kh} separated by distance $ql^2$. Since the particle-hole pair 
momentum is a good quantum number in the lowest Landau level, the density matrix can be expressed in 
terms of charge and pseudospin density operators 
\begin{equation}
\label{eq: mf5}
\hat{m}^{\mu}(\vec{q})= \frac{2\pi l^2}{A}\sum_{k_i\sigma_i}
\langle 0 k_1|e^{-i\vec{q}\cdot\hat{r}}|0 k_2\rangle 
c^{\dagger}_{k_1\sigma_1}\tau^{\mu}_{\sigma_1\sigma_2}c_{k_2\sigma_2}
\end{equation}
where $\tau^{\mu}$ are the Pauli matrices in the pseudospin space, and $|0k\rangle$ are the 
single-particle orbital eigenstates in the lowest Landau level ($n=0$). In the Hartree-Fock 
approximation, the bilayer single-particle Hamiltonian, when expressed in terms of charge and 
pseudospin density operators, becomes 
\begin{eqnarray}
\label{eq: mf6}
\frac{\hat{H}_{HF}}{A} & = & -\frac{\Delta_t}{4\pi l^2}\hat{m}^x_{q=0}-
	\frac{\Delta_v}{4\pi l^2}\hat{m}^z_{q=0} \nonumber \\
& + & \frac{1}{4\pi l^2}\sum_{\vec{p}}\left[\left(2v_0^{\vec{p}}-\Gamma_A^{\vec{p}}\right)
\langle\hat{m}^{0}_{\vec{p}}\rangle\hat{m}^0_{-\vec{p}}+
\left(2 v_x^{\vec{p}}-\Gamma_A^{\vec{p}}\right)\langle\hat{m}^{z}_{\vec{p}}\rangle\hat{m}^z_{-\vec{p}}+ 
\right]\nonumber \\
 & - & \frac{1}{4\pi l^2}\sum_{\vec{p}}\left[
\Gamma_E^{\vec{p}}\langle\hat{m}^{x}_{\vec{p}}\rangle\hat{m}^x_{-\vec{p}}+
\Gamma_E^{\vec{p}}\langle\hat{m}^{y}_{\vec{p}}\rangle\hat{m}^y_{-\vec{p}}
\right]
\end{eqnarray}
where $2\pi l^2v_\lambda(\vec{q})=V_\lambda(\vec{q}) e^{-q^2l^2/2}$ are the Coulomb interactions 
projected onto the lowest Landau level. 

In the mean-field approximation, the competition between single-particle terms and interaction 
contributions determines whether or not the system behaves as an easy-plane ferromagnet. Among 
single-particle terms interlayer tunneling amplitude $\Delta_t$ favors all pseudospins pointing along 
the positive $x$-axis whereas the bias-voltage $\Delta_v$ tends to align pseudospins along the 
positive $z$-axis leading to a charge-imbalance between the layers. Among the interaction 
contributions, the capacitive energy cost $v_x$ forces the pseudospins to lie in the $x$-$y$ plane 
whereas the exchange energy cost $\Gamma_E$ mandates their local alignment. The Hartree-Fock solution 
is a compromise between these competing tendencies and as a result the pseudospins lie in the $x$-$z$ 
plane. We consider uniform Hartree-Fock eigenstates parametrized by a {\it single} variational 
parameter, the polar angle $\theta$ in the $x$-$z$ plane in pseudospin space, corresponding to 
mean-field polarization $\langle\hat{m}_x(\vec{q})\rangle=\delta_{q0}\sin\theta$, 
$\langle\hat{m}_z(\vec{q})\rangle=\delta_{q0}\cos\theta$ and $\langle\hat{m}_y(\vec{q})\rangle=0$. 
In the basis of these eigenstates the Hartree-Fock Hamiltonian (\ref{eq: mf6}) is diagonal in 
intra-Landau level index and its structure in pseudospin-space is given by 
\begin{eqnarray}
\label{eq: mf7}
\hat{H}_{HF} & = & -h_z(\theta)\cdot\tau^z-h_x(\theta)\cdot\tau^x, \\
\label{eq: mf8}
h_z(\theta) & = & \frac{1}{2}\left(\Delta_v\cos\theta+\Delta_t\sin\theta\right)+\frac{\Gamma_E(0)}{2}-
\frac{\Delta_{vc}}{2}\cos^2\theta, \\
\label{eq: mf9}
h_x(\theta) & = & \frac{1}{2}\left(\Delta_v\sin\theta-\Delta_t\cos\theta\right)-
\frac{\Delta_{vc}}{2}\cos\theta\sin\theta.
\end{eqnarray}
Here we have defined layer-separation dependent interaction contribution to the bias-voltage~\cite{km} 
$-\Delta_{vc}(d)\cos\theta\equiv -2\left[v_x(0)-\Gamma_x(0)\right]\cos\theta$. The variational 
parameter $\theta$ is determined by the requirement that the mean-field Hamiltonian $\hat{H}_{HF}$ is 
diagonal {\it i. e.} that the total effective pseudospin field has the same orientation as the 
pseudospinors. 

Let us first concentrate on the case with zero interlayer tunneling amplitude, $\Delta_t=0$. In this 
case, the mean-field Hamiltonian is U(1)-invariant in the $x$-$y$ plane in the pseudospin space. The 
mean-field equation $h_x(\theta)=0$ has two solutions: $\theta_v=0$ and 
$\cos\theta_v=\Delta_v/\Delta_{vc}$. For small bias voltages, $\Delta_v\leq \Delta_{vc}$, the 
charge-unbalanced solution which spontaneously breaks the U(1) symmetry in the $x$-$y$ plane, 
$\cos\theta_v=\Delta_v/\Delta_{vc}$, is stable. The mean-field eigenstates are  
\begin{eqnarray}
\label{eq: mf10}
|\theta_v,+\rangle=\cos\frac{\theta_v}{2}|\uparrow\rangle+\sin\frac{\theta_v}{2}|\downarrow\rangle, \\
\label{eq: mf11}
|\theta_v,-\rangle=\sin\frac{\theta_v}{2}|\uparrow\rangle-\cos\frac{\theta_v}{2}|\downarrow\rangle, 
\end{eqnarray}
and the mean-field quasiparticle gap is $\Delta_{qp}=\Gamma_E(0)=\Delta_{sb}$. Note that surprisingly, 
the quasiparticle gap is independent of the bias-voltage $\Delta_v$.~\cite{brey,ch1} For 
$\Delta_v\leq\Delta_{vc}$ mean-field theory predicts that the charge-imbalance increases linearly 
with the applied bias-voltage, $m_z=\cos\theta_v=\Delta_v/\Delta_{vc}$, and that the $x$-$y$ component 
of the pseudospin polarization $m_x=\sin\theta_v=\sqrt{1-\Delta_v^2/\Delta_{vc}^2}$ decreases with the 
bias-voltage. The second solution of the mean-field equation $h_x(\theta)=0$, given by $\theta_v=0$, 
is the stable solution for large bias-voltages $\Delta_v\geq\Delta_{vc}$. In this case the layer-index 
is a good quantum number for the mean-field quasiparticles and the U(1) symmetry in the $x$-$y$ plane 
in pseudospin-space is unbroken. The mean-field eigenstates are single-particle states localized in 
the top and the bottom layer and the quasiparticle gap is given by
\begin{equation}
\label{eq: mf12}
\Delta_{qp}=\left[\Gamma_E(0)+\Delta_v-\Delta_{vc}\right]\geq \Delta_{sb}.
\end{equation}
The quasiparticle gap (\ref{eq: mf12}) represents a competition between the interlayer exchange-energy 
associated with interlayer phase-coherence, $\Gamma_E(0)$, and the capacitive energy 
$(\Delta_v-\Delta_{vc})$ associated with the charge-imbalance. In this mean-field state ($\theta=0$), 
all electrons occupy one layer, and the other layer is empty.~\cite{brey,ch1} 

\begin{figure}[t]
\begin{center}
\epsfxsize=4in
\epsffile{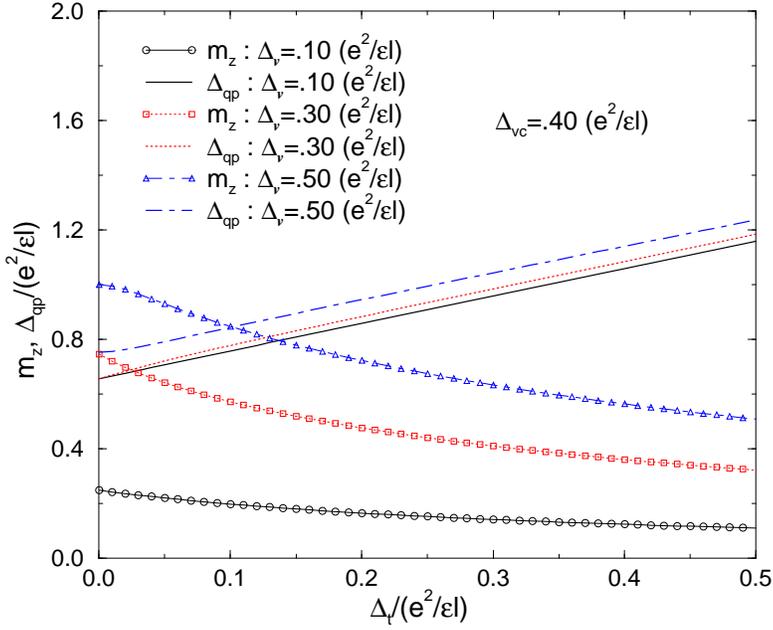}
\vspace{0.5cm}
\caption{The dimensionless charge-imbalance $m_z$ and the quasiparticle gap $\Delta_{qp}$ in the 
mean-field theory approximation. The layer separation is $d/l=1$. Note that at $\Delta_t=0$ the 
quasiparticle gap is independent of the bias-voltage for $\Delta_v<\Delta_{vc}$. The mean-field 
charge-imbalance decreases and the quasiparticle gap increases with tunneling amplitude $\Delta_t$.} 
\label{fig: 3dphase_mean_field}
\end{center}
\end{figure}

It is straightforward to extend this analysis to a system with nonzero tunneling amplitude $\Delta_t$ 
which explicitly breaks the U(1) symmetry in the pseudospin space. Because of the presence of two 
single-particle terms, the pseudospin always lies in the $x$-$z$ plane, but never entirely along the 
$x$-axis or the $z$-axis. In this case the solution to the equation $h_x(\theta)=0$ is obtained 
numerically. Figure~\ref{fig: 3dphase_mean_field} shows the mean-field charge-imbalance $m_z$ and 
quasiparticle gap $\Delta_{qp}$ as a function of tunneling amplitude $\Delta_t$ for various values of 
bias-voltage. At $\Delta_t=0$, we find that charge-imbalance is less than unity for 
$\Delta_v\leq\Delta_{vc}$ and the corresponding quasiparticle gaps are independent of the bias-voltage.
 As the tunneling amplitude increases we see that the charge-imbalance monotonically decreases and the 
quasiparticle gap $\Delta_{qp}$ increases.~\cite{brey} 

Now we consider collective excitations around the Hartree-Fock mean-field solution. Since the 
mean-field state is pseudospin polarized, collective excitations are modulations of the pseudospin 
field away from the mean-field configuration, {\it i. e.} pseudospin-waves. In bilayer systems, because 
of the difference between the interlayer and the intralayer Coulomb interactions, it is necessary to 
take into account fluctuations in direct and exchange channels to determine the spectrum of collective 
excitations.~\cite{haf} This spectrum is obtained most easily by using the auxiliary field functional 
integral approach.~\cite{kerman,ya1} In this approach we consider quadratic fluctuations of the 
order-parameter field around the Hartree-Fock state. The collective mode energies are obtained from 
zeros of the determinant of the quadratic fluctuation matrix. We can obtain the same results from the 
inverse generalized-random-phase-approximation susceptibility~\cite{haf}, or by considering quadratic 
expansion of the Hartree-Fock energy functional around the mean-field solution 
$\vec{m}(\vec{q})=\delta_{q0}(\sin\theta_v,0,\cos\theta_v)$, quantizing the fluctuations using bosonic 
(pseudospin-wave) creation and annihilation operators and diagonalizing the resulting pseudospin-wave 
Hamiltonian.~\cite{atm,ya2} The quadratic fluctuation matrix, or equivalently the inverse of the 
collective-mode propagator, is given by~\cite{ya1} 
\begin{equation}
\label{eq: mf13}
M(\vec{q},i\Omega)=\left[1+V^{A}(\vec{q})D(i\Omega)\right]
\end{equation}
where $\langle\alpha\beta|V^{A}|\gamma\delta\rangle=\langle\alpha\beta|V|\gamma\delta\rangle-
\langle\alpha\beta|V|\delta\gamma\rangle\equiv (V-V^{ex})$ is the antisymmetrized microscopic 
interaction, $D$ is the Hartree-Fock susceptibility matrix and we have suppressed the pseudospin 
pair-labels. When represented in the basis of mean-field eigenstates $|\theta_v,\pm\rangle$, the 
fluctuation matrix becomes 
\begin{eqnarray}
\label{eq: mf14}
\left[1+V^A(\vec{q})D(i\Omega_n)\right] & = & 
\left[\begin{array}{cc}
1+ \left(A-\Gamma_E\right)\cdot\frac{1}{i\Omega_n+\Delta_{qp}} & 
-A\cdot\frac{1}{i\Omega_n-\Delta_{qp}} \\ 
A\cdot\frac{1}{i\Omega_n+\Delta_{qp}} & 
1-\left(A-\Gamma_E\right)\cdot\frac{1}{i\Omega_n-\Delta_{qp}}
\end{array} \right] 
\end{eqnarray}
where $A(\vec{q},\theta_v)=\left[v_x(\vec{q})-\Gamma_x(\vec{q})\right]\sin^2\theta_v$. In diagrammatic 
language the exchange contributions, $\Gamma_E(\vec{q})$ and $\Gamma_x(\vec{q})\sin^2\theta_v$, arise 
from summing the particle-hole ladders whereas the Hartree contribution $v_x(\vec{q})\sin^2\theta_v$ 
arises from the particle-hole bubble sum.~\cite{ya1} Notice that the term $A(\vec{q},\theta_v)$ 
involves only the {\it anisotropic} part of the Coulomb interaction. We obtain the following simple 
expression for the dispersion of pseudospin-waves, $i\Omega_n=E_{sw}(\vec{q},\theta_v)$, from the 
zeros of the fluctuation matrix
\begin{eqnarray}
\label{eq: mf15}
E^2_{sw}(\vec{q},\theta_v)& = & a_{\theta_v}(\vec{q})\cdot b_{\theta_v}(\vec{q}), \\
\label{eq: mf16}
a_{\theta_v}(\vec{q}) & = & \Delta_{qp}-\Gamma_E(\vec{q})+2A(\vec{q},\theta_v), \\
\label{eq: mf17}
b_{\theta_v}(\vec{q}) & = & \Delta_{qp}-\Gamma_E(\vec{q}).
\end{eqnarray}
The two energy terms in Eq.(\ref{eq: mf15}) have simple interpretations. The first term, 
$a_{\theta_v}(\vec{q})$, is the energy cost associated with charge-imbalance fluctuations. The second 
term, $b_{\theta_v}(\vec{q})$, denotes the energy cost associated with phase-fluctuations in the 
$x$-$y$ plane. Eq.(\ref{eq: mf15}) reproduces the pseudospin-wave dispersion for the phase-coherent 
state~\cite{haf,apb} when the bias-voltage is zero or, equivalently, when $\theta_v=\pi/2$. For zero 
interlayer tunneling and small bias-voltages, the quasiparticle gap is $\Delta_{qp}=\Gamma_E(0)$. Then 
it is easy to see that the cost for uniform phase-fluctuations is zero, $b_{\theta_v}(\vec{q})=0$, 
consistent with Goldstone's theorem. On the other hand, for $\Delta_v>\Delta_{vc}$ the mean-field 
solution does not break the U(1) symmetry ($\theta_v=0$) and the pseudospin-wave dispersion simplifies 
to $E_{sw}(\vec{q})=\Delta_{qp}-\Gamma_E(\vec{q})$. 

\begin{figure}[t]
\begin{center}
\begin{minipage}{20cm}
\begin{minipage}{9cm}
\epsfxsize=3.3in
\epsffile{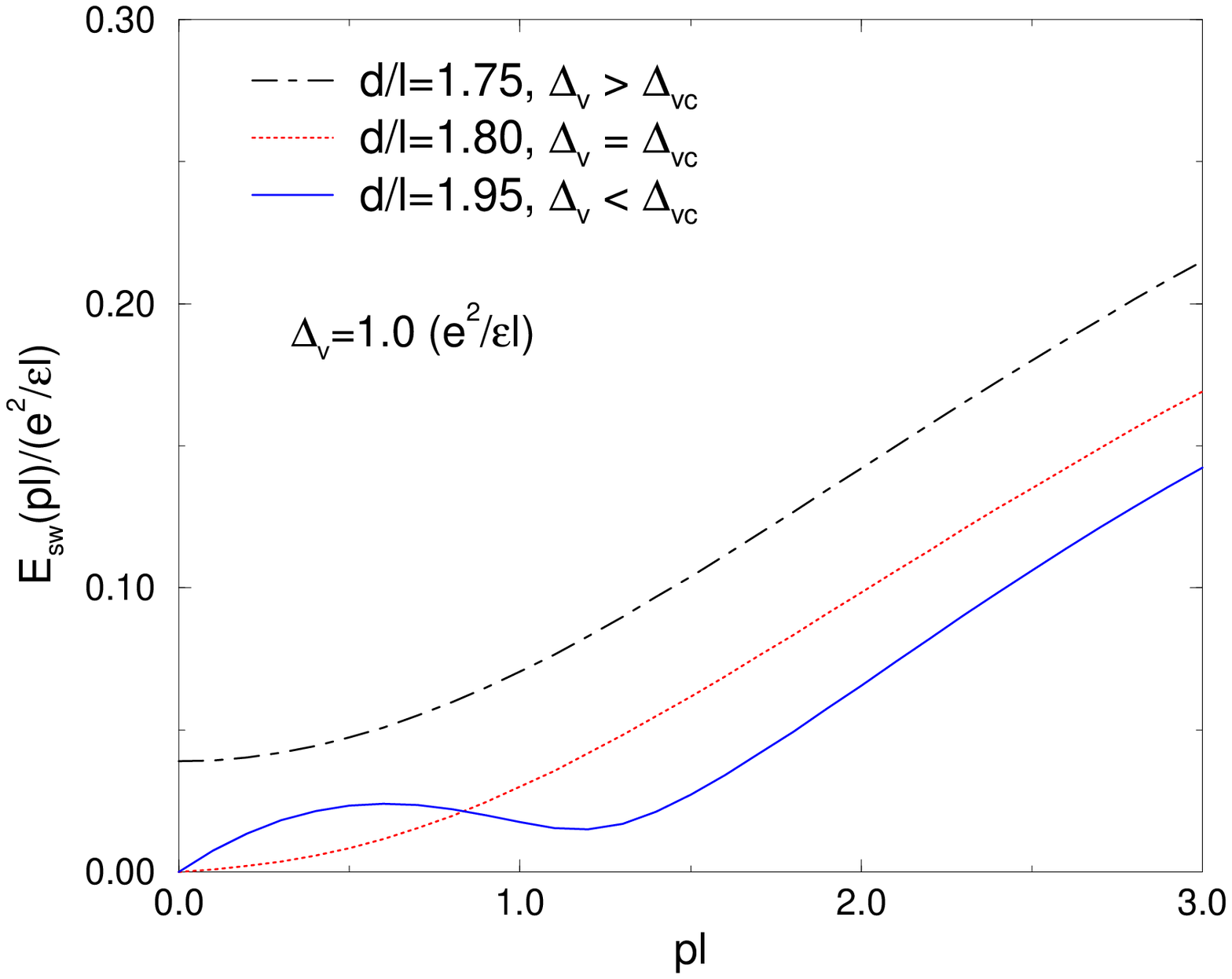}
\end{minipage}
\begin{minipage}{9cm}
\epsfxsize=3.3in
\epsffile{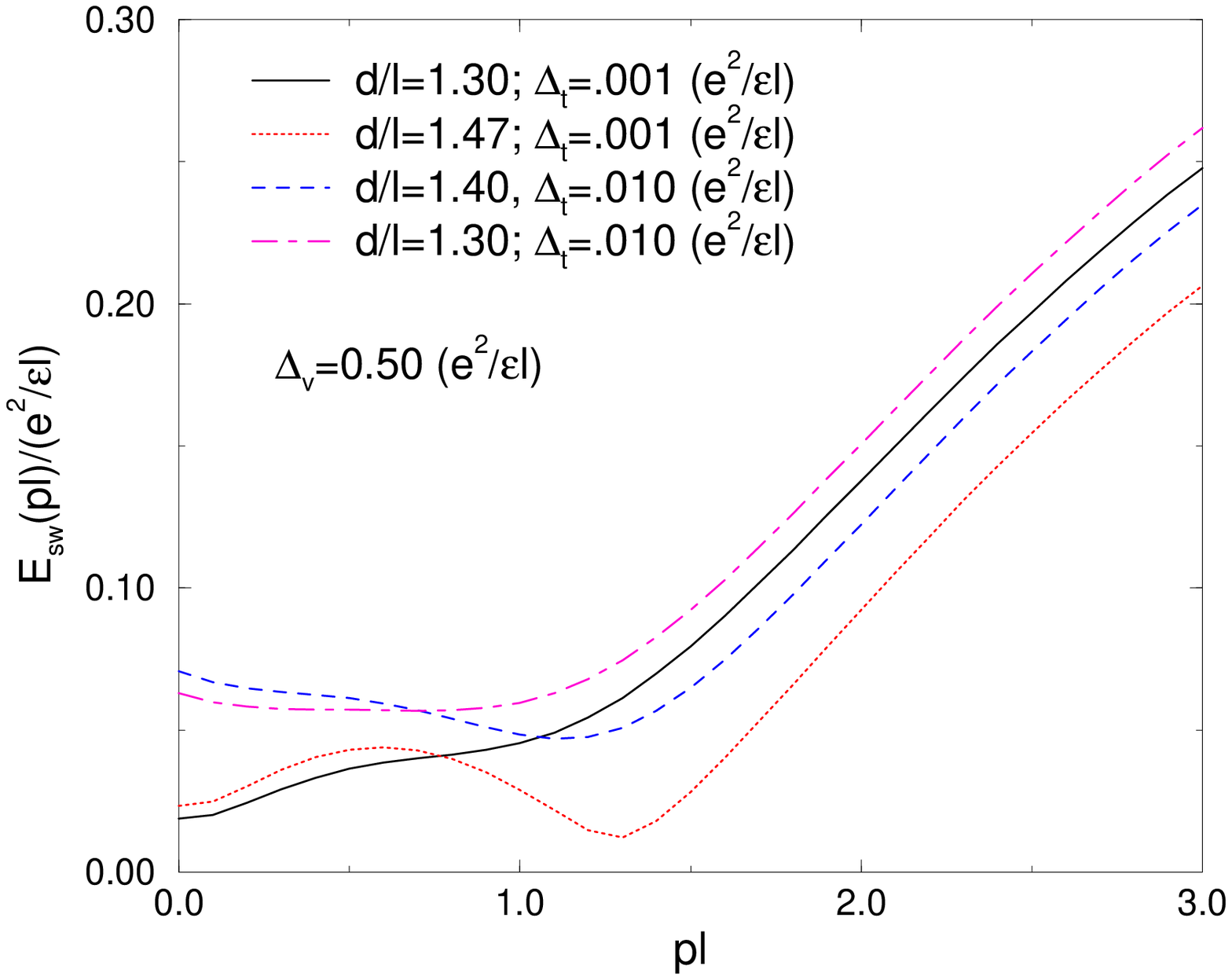}
\end{minipage}
\end{minipage}
\caption{Typical generalized RPA collective mode dispersions for $\Delta_t=0$ (left) and 
$\Delta_t\neq 0$ (right). When $\Delta_t=0$, the pseudospin waves are gapless and have a linear 
dispersion for $\Delta_v<\Delta_{vc}$ whereas for $\Delta_v>\Delta_{vc}$, the mode is gapped and has a 
quadratic dispersion. When $\Delta_t\neq 0$, the pseudospin waves are gapped at zero wave-vector. The 
non-monotonic behavior of the dispersion arises from competing Hartree and exchange fluctuations which 
lead to the softening of pseudospin-waves near $pl\approx 1$ for sufficiently large $d$.}
\label{fig: dispersion}
\end{center}
\end{figure}

Figure~\ref{fig: dispersion} shows typical dispersions for the various cases discussed above. When 
$\Delta_t=0$ (left figure), for $\Delta_v<\Delta_{vc}$, the mean-field state breaks the symmetry in 
the $x$-$y$ plane spontaneously and the resulting pseudospin-waves are gapless and {\it linearly} 
dispersing at long wavelengths. At $\Delta_v=\Delta_{vc}$ the Gaussian fluctuation Hamiltonian is 
isotropic in pseudospin-space. Therefore the collective excitations are gapless, but have a 
{\it quadratic} dispersion. In contrast, when $\Delta_v>\Delta_{vc}$ the mean-field solution does not 
break the U(1) symmetry in the plane and consequently the pseudospin waves are gapped at zero 
wave-vector. When the interlayer tunneling is nonzero (right figure) we see that the pseudospin waves 
are always gapped at zero wave-vector. In all cases, the non-monotonic behavior of the dispersion near 
$ql\approx 1$ arises because of the competing energy costs from Hartree and exchange fluctuations in 
the term $a_{\theta_v}(\vec{q})$. We emphasize that since the functional integral approach expresses 
the properties of collective excitations only in terms of the mean-field quasiparticle gaps 
$\Delta_{qp}$ and the microscopic interaction $V$, it is possible to obtain simple analytical 
expressions for collective-mode spectra in different cases with relative ease. 

%------------------------------------------------------------------------------------------------------%

\section{Phase transition in biased bilayer systems}
\label{sec: phase}

In this section we will explore the phase-diagram of a bilayer system in the $(d,\Delta_v)$ plane as 
a function of interlayer tunneling. In the preceding section we considered {\it uniform} mean-field 
solutions and collective excitations around them for a bilayer system in the presence of interlayer 
tunneling and bias-voltage. The dispersion of pseudospin-waves in the phase-coherent regime, 
Eq.(\ref{eq: mf16}), implies that for a given set of parameters $(\Delta_t,\Delta_v)$ the energy cost 
associated with charge-imbalance fluctuations, $a_{\theta_v}(\vec{q})$, vanishes near $ql\approx 1$ at 
critical layer separation $d_{cr}(\Delta_t,\Delta_v)$. This softening of the collective mode at finite 
wave-vector indicates the propensity toward the formation of pseudospin-density waves with wave-vector 
$ql\approx 1$, leading, in this theory, to the collapse of the quasiparticle gap and the disappearance 
of the $\nu=1$ quantum Hall effect. We use the softening of pseudospin-waves as the criterion to 
determine the phase-boundary between $\nu=1$ quantum Hall regime and the compressible 
regime.~\cite{apb,ch1} In the Hartree-Fock approximation the state on the large $d$ side of this 
transition is~\cite{cote} a charge density wave state, which we believe is quantum melted yielding a 
compressible state. 

\begin{figure}[h]
\begin{center}
\epsfxsize=4in
\epsffile{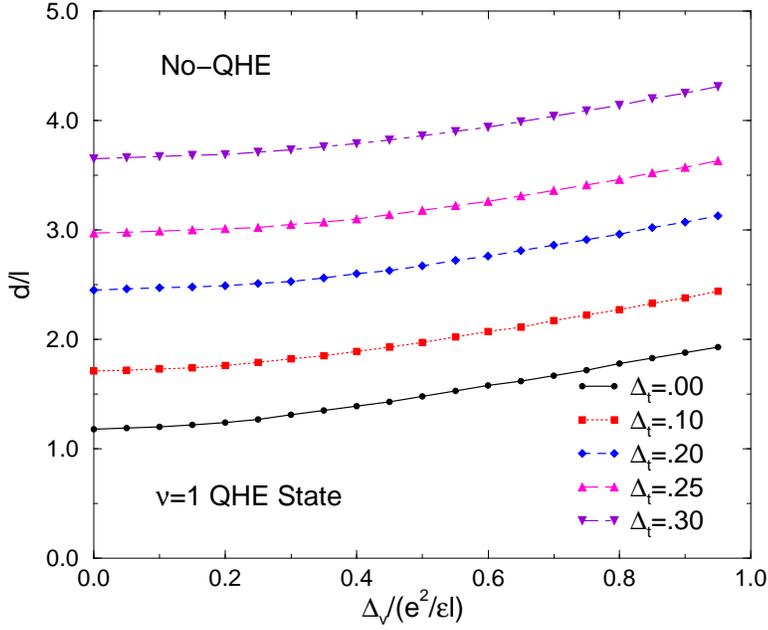}
\vspace{0.5cm}
\caption{Phase-boundary between the compressible and the $\nu=1$ quantum Hall state in the 
$(d,\Delta_v)$ plane. This phase-diagram predicts that as the interlayer bias-voltage is increased, 
incompressible states with (spontaneous) interlayer phase-coherence survive to larger layer 
separations.}
\label{fig: 3dphase_diagram}
\end{center}
\end{figure}

Figure~\ref{fig: 3dphase_diagram} shows the phase-boundary in the $(d,\Delta_v)$ plane which separates 
the $\nu=1$ quantum Hall state from the compressible state for various values of interlayer tunneling 
amplitude. We see that the phase diagrams are qualitatively similar for different tunneling 
amplitudes. {\it This phase-diagram predicts that starting from a compressible state, a bilayer system 
at total filling factor $\nu=1$ will develop (spontaneous) interlayer phase-coherence and consequently 
show the $\nu=1$ quantum Hall effect when an appropriate bias-voltage is applied.} This is the first 
principle conclusion of this work and results from the property that pseudospin-waves around the 
$\nu=1$ quantum Hall state are stiffened by the bias-voltage.~\cite{ch1,arh,sawada,ez} The emergence 
of the $\nu=1$ quantum Hall plateau had been predicted by Brey {\it et al.}~\cite{brey} when the 
applied bias-voltage exceeds the critical bias-voltage, $\Delta_v\geq\Delta_{vc}$, and all charge is 
in one layer. On the other hand, our calculations predict the emergence of the $\nu=1$ plateau for 
$\Delta_v<\Delta_{vc}$. This entire 
phase-diagram has not yet been explored experimentally; in fact most experiments have been confined to 
zero bias-voltage, $\Delta_v=0$.~\cite{sqm,spi1,spi2} In this limit the $y$-intercepts of 
phase-boundaries for various tunneling amplitudes reproduce the phase-diagram in the $(d,\Delta_t)$ 
plane~\cite{sqm} which had been obtained previously.~\cite{apb} It is worthwhile to point out that we 
consider bilayer systems with finite electrical potential difference $\Delta_v$, but with zero 
electrochemical potential difference, which results in a finite charge-imbalance and zero interlayer 
current in an equilibrium situation. The experimental investigation of the bias-voltage driven 
phase-transition we predict is feasible and would partially confirm that the weak coupling description 
of the coherent state is the correct description.

%------------------------------------------------------------------------------------------------------%

\section{Renormalization of Effective Field Theory Parameters}
\label{sec: renorm}

In the quantum Hall regime, the low energy dynamics of the charge-unbalanced and phase-coherent state 
is described an effective pseudospin field theory in which the Hamiltonian is parametrized by an 
anisotropy mass $\beta$ which represents the cost of uniform charge-imbalance fluctuations, and 
pseudospin stiffness $\rho_s$ which expresses the energy cost associated with 
phase-fluctuations.~\cite{km,ky} These parameters should be evaluated from the microscopic 
Hamiltonian, Eqs.(\ref{eq: mf1}), (\ref{eq: mf2}). In this section we will calculate the low-energy 
parameters by using approximations for the ground state energy which take into account the effect of 
collective excitations around the Hartree-Fock mean-field state, {\it i.e.} using the generalized 
random phase approximation (GRPA).

We first briefly recall the general expressions for the GRPA energy around a Hartree-Fock 
mean-field.~\cite{kerman,ya1} For an interacting system with a kinetic term $K\hat{\rho}$ and a 
two-body interaction $V$, the mean-field approximation for the ground state energy is given by  
\begin{equation}
\label{eq: renorm1}
E_{qp}=\sum_{h}K_{hh}+\frac{1}{2}\sum_{hh'}V^{A}_{hh'hh'}
\end{equation}
where the labels $h,h'$ stand for the occupied (hole) states. Thus $E_{qp}$ is just the expectation 
value of the microscopic Hamiltonian in the Hartree-Fock mean-field ground state. The correlation 
energy contribution $E_c$ which takes into account the effects of collective excitations around the 
Hartree-Fock mean-field state is given by 
\begin{equation}
\label{eq: renorm2}
E_c=\frac{1}{2}\sum_{\omega_\nu>0}\omega_\nu-\frac{1}{2}\sum_{p,h}\left(e_p-e_h\right)
	-\frac{1}{2}\sum_{p,h}V^{A}_{hpph}.
\end{equation}
Here $\omega_\nu$ are the energies of collective modes around the Hartree-Fock mean-field, $\epsilon_a$ 
denote the energies of the Hartree-Fock quasiparticles, and $p,h$ stand for the unoccupied (particle) 
and occupied (hole) states, respectively. We obtain renormalized (mean-field) parameters by using the 
approximation $E_G=E_{qp}+E_c$ ($E_G=E_{qp}$). For bilayer systems since the particle-hole pair 
momentum is a good quantum number~\cite{kh} it is straightforward to obtain explicit expressions for 
the mean-field and correlation contributions.~\cite{ya1}

We calculate the dimensionless charge-imbalance by evaluating the bias-dependence of the ground state 
energy  
\begin{equation}
\label{eq: renorm3}
m_z(\Delta_v) =  -\frac{4\pi l^2}{A}\frac{\partial E_G(\Delta_v)}{\partial\Delta_v}.
\end{equation}
The anisotropy mass is obtained from the bias-dependence of charge-imbalance, 
$\beta^{-1}=8\pi l^2\partial m_z/\partial\Delta_v$. For simplicity, we consider a system with zero 
interlayer tunneling amplitude. In the phase-coherent state Eq.(\ref{eq: renorm1}) for the ground 
state energy gives 
\begin{equation}
\label{eq: renorm4}
\frac{E_{qp}}{A}=-\frac{1}{4\pi l^2}\left[\frac{\Delta_v^2}{2\Delta_{vc}}+\Gamma_0(0)\right]. 
\end{equation}
We recover the mean-field result for the charge-imbalance $m_z^{HF}=\Delta_v/\Delta_{vc}$ and the 
anisotropy mass, $\beta_{HF}=\Delta_{vc}/8\pi l^2$ by using this mean-field approximation 
(\ref{eq: renorm4}) for the ground state energy.~\cite{km,ky} The correlation energy contribution due 
to the pseudospin-waves is given by 
\begin{equation}
\label{eq: renorm5} 
\frac{E_c}{A}=\frac{1}{2A}\sum_{\vec{p}}\left[E_{sw}(\vec{p},\theta_v)-\Delta_{sb}\right]-
\frac{1}{4\pi l^2}\left[\frac{\Delta_v^2}{2\Delta_{vc}}+\Gamma_x(0)\right].
\end{equation}
Using explicit expressions for the collective-mode dispersion (\ref{eq: mf15}) we get the following 
simple analytical expression for the renormalized charge-imbalance
\begin{equation}
\label{eq: renorm6}
m_z  = 2m_z^{HF}-\frac{2\pi l^2}{A}\sum_{\vec{p}}\left[\frac{b_p}{2E_{sw}(p)}
\frac{\partial a_p}{\partial\Delta_v}\right].
\end{equation}
Eq.(\ref{eq: renorm6}) is the second main result of this work. The correlation term in 
Eq.(\ref{eq: renorm6}) gives a large negative contribution, which leads to a strong suppression of 
charge-imbalance $m_z$ near the phase-boundary $d_{cr}$. The correlation energy is larger in magnitude 
at small $m_z$ and therefore favors smaller values of $m_z$. It is easy to obtain an explicit 
expression for renormalized anisotropy mass from Eq.(\ref{eq: renorm6}), though it is not particularly 
illuminating. We note that the correlation energy contribution (\ref{eq: renorm5}) diverges because of 
the long range of Coulomb interaction, but the renormalized observables are finite and well-defined.

\begin{figure}[h]
\begin{center}
\epsfxsize=4in
\epsffile{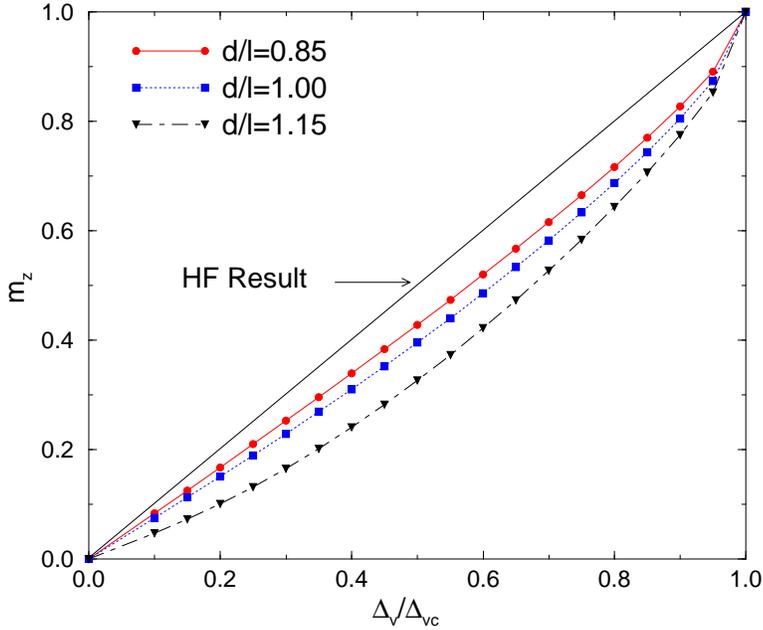}
\caption{Renormalized charge-imbalance $m_z$ for various layer separations at $\Delta_t=0$. The 
mean-field result is $m_z^{HF}=\Delta_v/\Delta_{vc}$. The absence of renormalization at $\Delta_v=0$ 
and at $\Delta_v=\Delta_{vc}$ is expected from symmetry considerations. As $d\rightarrow d_{cr}$, the 
collective fluctuations soften at a finite wave-vector leading to a substantial renormalization of the 
interlayer capacitance $\beta^{-1}=8\pi l^2 \partial m_z/\partial\Delta_v$.} 
\label{fig: renormalized_mz}
\end{center}
\end{figure}

Figure~\ref{fig: renormalized_mz} compares our results for the $\Delta_v$ dependence of the mean-field 
and renormalized charge-imbalance $m_z$. For $\Delta_v>\Delta_{vc}$, the mean-field solution does not 
break the U(1) symmetry in the $x$-$y$ plane in pseudospin-space. Therefore, for 
$\Delta_v\geq \Delta_{vc}$ collective excitations do not renormalize the mean-field result, $m_z=1$.
We see that for $\Delta_v<\Delta_{vc}$, as the layer separation $d$ increases the charge-imbalance is 
strongly renormalized because of the imminent softening of pseudospin-waves. Thus the renormalized 
charge-imbalance (\ref{eq: renorm6}) reflects the softening of the pseudospin-waves at finite wave 
vector. It is clear from Fig.~\ref{fig: renormalized_mz} that fluctuations enhance the anisotropy mass 
$\beta$ or equivalently reduce the interlayer capacitance, $8\pi l^2\partial m_z/\partial\Delta_v$, at 
small bias-voltages, whereas at large bias voltages, the interlayer capacitance is increased by 
collective fluctuations. Thus our analysis predicts that at near critical layer separation, the 
interlayer capacitance - {\it an experimentally measurable quantity} - will be strongly renormalized 
from its mean-field value. 

Now we turn to the calculation of bias-dependent pseudospin stiffness $\rho_s(\Delta_v)$. In the 
absence of external tunneling the Hartree-Fock equations support spiral-state solutions with pairing 
wave vector $\vec{Q}=Q\hat{x}$ where the mean-field polarization varies linearly with in-plane 
coordinate, $\vec{m}=(\cos Qx\sin\theta_v,\sin Qx\sin\theta_v,\cos\theta_v)$. The pseudospin stiffness 
is obtained from the dependence of the ground state energy $E_G(\Delta_v,Q)$ on the spiral wave vector 
\begin{equation}
\label{eq: renorm7}
\rho_s(\Delta_v)=\frac{1}{A}\frac{\partial^2 E_G(\Delta_v,Q)}{\partial Q^2}|_{Q=0}.
\end{equation}

For spiral states in the presence of a bias-voltage, the mean-field eigenstates are given by 
\begin{eqnarray}
\label{eq: renorm8}
|+,k,\theta_v\rangle & = &\cos\frac{\theta_v}{2}|\uparrow\rangle+
\sin\frac{\theta_v}{2}e^{iQkl^2}|\downarrow\rangle, \\
\label{eq: renorm9}
|-,k,\theta_v\rangle & = &\sin\frac{\theta_v}{2}|\uparrow\rangle-
\cos\frac{\theta_v}{2}e^{iQkl^2}|\downarrow\rangle,
\end{eqnarray}
where $\cos\theta_v=\Delta_v/\Delta_{vc}^Q$ determines the angle $\theta_v$ and 
$\Delta_{vc}^Q(d)=\left[2v_x(0)-\Gamma_A(0)+\Gamma_E(Q)\right]$ denotes the interaction contribution 
to the interlayer potential difference. The quasiparticle gap in the phase-coherent state is 
$\Delta_{qp}=\Delta_{Q}=\Gamma_E(Q)$. Using the functional integral approach it is straightforward to 
obtain the dispersion of pseudospin-waves, 
\begin{equation}
\label{eq: renorm10}
E_{sw}^2(\vec{q},Q,\theta_v) = a(\vec{q},Q,\theta_v)\cdot b(\vec{q},Q,\theta_v).
\end{equation}
Here the cost of charge-imbalance fluctuations is given by 
\begin{eqnarray}
\label{eq: renorm11}
a(\vec{q},Q,\theta_v) & = & \Delta_{Q}+\left[2v_x^{Q}(\vec{q})-\Gamma_A(\vec{q})\right]\sin^2\theta_v - 
\frac{\cos\theta_v}{2}\left[\Gamma_E(\vec{q}-Q\hat{x})-\Gamma_E(\vec{q}+Q\hat{x})\right] \nonumber\\
& - & \frac{\cos^2\theta_v}{2}\left[\Gamma_E(\vec{q}-Q\hat{x})+\Gamma_E(\vec{q}+Q\hat{x})\right]
\end{eqnarray}
with $2\pi l^2v_x^Q(\vec{q})=e^{-q^2l^2/2}\left[V_A(\vec{q})-V_E(\vec{q})\cos Qq_yl^2\right]/2$. On 
the other hand the energy cost of phase-fluctuations is 
\begin{equation}
\label{eq: renorm12}
b(\vec{q},Q,\theta_v) = \Delta_{Q}-\frac{1}{2}\left[(1+\cos\theta_v)\Gamma_E(\vec{q}-Q\hat{x})+ 
(1-\cos\theta_v)\Gamma_E(\vec{q}+Q\hat{x})\right].
\end{equation}
The qualitative behavior of the dispersion (\ref{eq: renorm10}) is similar to the dispersion of 
pseudospin-waves in a uniform phase-coherent state (Fig.~\ref{fig: dispersion}); the collective 
excitations are linearly dispersing at long wavelength and have a local minimum near $ql\approx 1$ for 
large layer-separations. It is clear from Eqs.(\ref{eq: renorm12}) and (\ref{eq: renorm10}) that the 
cost of global phase rotation in the $x$-$y$ plane in pseudospin-space is zero and that the collective 
mode has vanishing energy at long wavelength. We emphasize that the anisotropic nature of the 
dispersion is related to the choice of the pairing wave vector $\vec{Q}=Q\hat{x}$, and the 
non-monotonic behavior of the dispersion is a result of competing Hartree and exchange contributions 
in Eq.(\ref{eq: renorm11}). The pseudospin-wave dispersion (\ref{eq: renorm10}) reproduces the earlier 
result, Eq.(\ref{eq: mf15}), when $Q=0$ and the results for supercurrent dispersion in the absence of 
bias voltage.~\cite{ya1,kmoonprl} 

\begin{figure}[h]
\begin{center}
\epsfxsize=4in
\epsffile{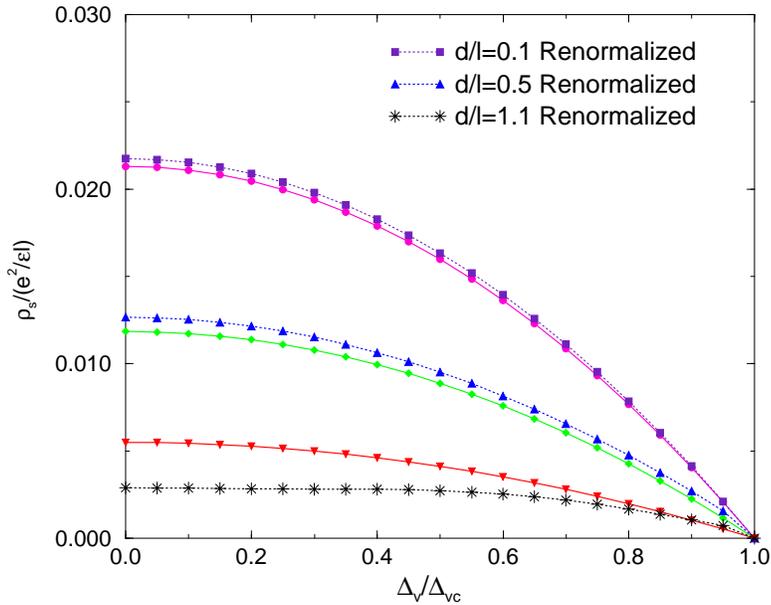}
\caption{Mean-field (solid lines) and renormalized (dotted lines) pseudospin-stiffness 
$\rho_s(\Delta_v)$. The interlayer tunneling amplitude $\Delta_t=0$. For $d/l< 1$ the 
pseudospin-stiffness is {\it enhanced} by collective fluctuations. When the layer separation is large, 
$d/l>1$, the stiffness is substantially renormalized at small bias-voltage; large bias-voltage 
stabilizes the collective mode and consequently leads to small renormalization of pseudospin 
stiffness. The intercepts on $y$-axis for various layer separations reproduce results for a balanced 
bilayer system with spontaneous interlayer phase-coherence.~\cite{ya1}} 
\label{fig: renorm_rho}
\end{center}
\end{figure}

In the present case, the mean-field ground state energy contribution is 
\begin{equation}
\label{eq: renorm13}
\frac{E_{qp}}{A}=-\frac{1}{4\pi l^2}\left[\frac{\Delta_v^2}{2\Delta_{vc}^Q}+\frac{\Delta_Q}{2}
	+\frac{\Gamma_A(0)}{2}\right],
\end{equation}
and the mean-field bias-dependent pseudospin stiffness~\cite{dns,ch1} is given by 
$\rho_s(\Delta_v)=\rho_{HF}\sin^2\theta_v=\rho_{HF}(1-\Delta_v^2/\Delta_{vc}^2)$ where we have used 
the fact that $\partial^2\Delta_Q/\partial Q^2=-8\pi l^2\rho_{HF}$. The correlation contribution to 
the ground state energy is 
\begin{equation}
\label{eq: renorm14}
\frac{E_c}{A}=\frac{1}{2A}\sum_{\vec{p}}\left[E_{sw}(\vec{p},Q,\theta_v)-\Delta_Q\right]-
	\frac{1}{4\pi l^2}\left[\frac{\Delta_v^2}{2\Delta_{vc}^Q}-\frac{\Delta_Q}{2}
	+\frac{\Gamma_A(0)}{2}\right].
\end{equation}
We obtain renormalized pseudospin stiffness from the approximate ground state energy which includes 
the effects of collective quantum fluctuations, $E_G=E_{qp}+E_c$, 
\begin{equation}
\label{eq: renorm15}
\rho_s(\Delta_v)=\frac{1}{2A}\sum_{\vec{p}}\left[\frac{\partial ^2E_{sw}}
	{\partial Q^2}+8\pi l^2\rho_{HF}\right]-2\rho_{HF}\frac{\Delta_v^2}{\Delta_{vc}^2}.
\end{equation}
This remarkably simple expression for renormalized stiffness is the third major result of this work. 
Figure~\ref{fig: renorm_rho} shows the bias-dependent mean-field (solid) and renormalized (dotted) 
stiffness as a function of charge-imbalance for various layer separations. At small layer separations 
$d/l< 1$ the stiffness is not strongly renormalized although it is {\it enhanced} by collective 
fluctuations.~\cite{ya1} However, when $d\geq l$, the pseudospin stiffness is substantially 
renormalized  from its mean-field value.~\cite{dns} We see that the difference between the mean-field 
and the renormalized stiffness decreases as the bias-voltage increases, which is consistent with the 
fact that the bias-voltage stabilizes the collective excitations.~\cite{ch1,arh,sawada,ez}

%------------------------------------------------------------------------------------------------------%

\section{Summary}
\label{sec: summary}

We have considered the effect of interlayer bias-voltage on bilayer quantum Hall systems at total 
filling factor $\nu=1$ using the Hartree-Fock mean-field theory and we have used the auxiliary field 
functional integral approach to study the collective excitations around the mean-field. 

Our stability analysis of the mean-field state predicts that bilayer quantum Hall systems can be driven 
from compressible to incompressible coherent states by the application of a bias voltage. This is the 
central result of the present work. We remind the reader that we consider equilibrium states with 
nonzero charge-imbalance but zero interlayer current; in other words, there is no electrochemical 
potential difference between the two layers, although there is an applied electrical potential 
difference $\Delta_v$. The experimental verification (or falsification) of this phase-transition seems 
feasible; indeed we hope that the results presented here will provide further motivation for 
experiments which explore the phase-diagram of a bilayer system in the parameter space 
$(d,\Delta_v,\Delta_t)$. In our microscopic theory, the coherent state is stabilized by the interlayer 
correlations it establishes which yield an interlayer exchange energy. The formation of an 
incompressible state weakens correlations within each layer, compared to the case of isolated layers. 
The competition between coherent and incoherent states is therefore one between interlayer and 
intralayer correlation energies. Our finding that at fixed layer separation a bias voltage favors 
coherent states reflects the lesser importance of intralayer correlations when the filling factor 
within that layer moves to either side of $\nu=1/2$. Verification of this prediction would add to 
earlier experimental evidence that supports the quantitative reliability of weak coupling 
considerations in describing coherent bilayer systems. 

In bilayer systems, the effect of quantum fluctuations on the mean-field ground state can be tuned 
over a large range by simply changing the ratio of layer separation and typical inter-particle spacing. 
We have presented simple analytical expressions for the renormalized charge-imbalance, 
Eq.(\ref{eq: renorm6}), and the bias-dependent pseudospin-stiffness, Eq.(\ref{eq: renorm15}), which 
take into account the effect of these collective quantum fluctuations around the Hartree-Fock state. 
These expressions reflect the rapidly increasing strength of collective fluctuations near the critical 
layer separation $d_{cr}$ and subsequently show a strong suppression of the charge-imbalance and the 
pseudospin-stiffness near the phase-boundary. Our analysis predicts that the interlayer capacitance 
is strongly renormalized from its mean-field value at layer separations $d\approx l$. In the present 
treatment we have restricted ourselves to Gaussian fluctuations, or equivalently, non-interacting 
collective excitations. Although our treatment of fluctuations is not necessarily accurate in the 
critical region close to the phase-boundary, we expect the results to be qualitatively similar.

In this paper, on several occasions, we have restricted ourselves to zero interlayer tunneling 
amplitude. It is straightforward to extend the analysis to finite tunneling amplitude. For example, we 
have presented spiral states are metastable solutions of the Hartree-Fock equations when the 
interlayer tunneling amplitude is zero. These spiral states are also realized as mean-field ground 
states in bilayer systems with nonzero tunneling amplitude in the presence of an in-plane magnetic 
field $B_{||}$.~\cite{sqm} The analysis presented here remains valid when the quasiparticle gaps and 
collective-mode spectra are evaluated in the presence of interlayer tunneling. 

In the present analysis we have assumed, implicitly, that the steady-state filling factors $\nu_T$ and 
$\nu_B$ do not correspond to quantum Hall states for isolated layers. If the filling factors 
correspond to such fractions {\it e. g.} $\nu_T=1/3$ and $\nu_B=2/3$, at large layer separations, the 
compressible phase suggested by softening of pseudospin-waves will be usurped by the $\nu_T=1/3$ and 
$\nu_B=2/3$ states which are strongly correlated within each layer. 

%------------------------------------------------------------------------------------------------------%

\section*{acknowledgments}
It is a pleasure to thank Jim Eisenstein, Charles Hanna and Mansour Shayegan for discussions, 
suggestions and comments. After this work was completed, we became aware of related work in bilayer 
systems (in the absence and presence of magnetic field) by Charles Hanna and co-workers~\cite{ch1,ch2} 
which did not, however, consider correlation effects. This work was supported by Robert A. Welch 
foundation and by NSF grant No. DMR0115947.

%------------------------------------------------------------------------------------------------------%

\newpage

%\newpage

%------------------------------------------------------------------------------------------------------%

%======================================================================================================%

\end{document}